\documentclass[manuscript,screen,acmsmall]{acmart}

\begin{document}

\title{Human Perception of LLM-generated Text Content in Social Media Environments}

\author{Kristina Radivojevic}
\email{kradivo2@nd.edu}
\orcid{0000-0002-1645-5945}
\authornotemark[1]
\affiliation{%
  \institution{University of Notre Dame}
  \city{Notre Dame}
  \state{Indiana}
  \country{USA}
}

\author{Matthew Chou}
\affiliation{%
    \institution{University of Notre Dame}
    \city{Notre Dame}
    \state{Indiana}
    \country{USA}}
  \email{mchou3@nd.edu}

\author{Karla Badillo-Urquiola}
\affiliation{%
   \institution{University of Notre Dame}
   \city{Notre Dame}
   \state{Indiana}
   \country{USA}}

\author{Paul Brenner}
\affiliation{%
    \institution{University of Notre Dame}
    \city{Notre Dame}
    \state{Indiana}
    \country{USA}}

\begin{abstract}
  Emerging technologies, particularly artificial intelligence (AI), and more specifically Large Language Models (LLMs) have provided malicious actors with powerful tools for manipulating digital discourse. LLMs have the potential to affect traditional forms of democratic engagements, such as voter choice, government surveys, or even online communication with regulators; since bots are capable of producing large quantities of credible text. To investigate the human perception of LLM-generated content, we recruited over 1,000 participants who then tried to differentiate bot from human posts in social media discussion threads. We found that humans perform poorly at identifying the true nature of user posts on social media. We also found patterns in how humans identify LLM-generated text content in social media discourse. Finally, we observed the Uncanny Valley effect in text dialogue in both user perception and identification. This indicates that despite humans being poor at the identification process, they can still sense discomfort when reading LLM-generated content. 
\end{abstract}




\maketitle

\section{Introduction}

Decentralization of information empowers individuals with access to technology to control, influence, and shape narratives. This shift was previously seen with social media platforms that have allowed greater access to disparate and previously inaccessible audiences, challenging the pre-established status-quo, and creating new types of celebrities and industries. The dissemination of highly sophisticated propaganda and disinformation is no longer confined to nation-states or highly technical organizations. Internet users can have an impact on a global scale, with rapid proliferation across multiple channels, making identification and analysis of content more challenging, especially in a manner rapid enough to counter intended effects. Because of that, risk to reputation and policy is greatly impacted on a global scale. Although there has always been unreliable or sometimes fake information on the Internet, users often choose to trust what they see and hear in discourse on their preferred social media channels. A rise in AI and LLM technologies posing as humans may contribute to the normalization of unreliable information. There are many types of manipulation in digital media, and while the most popular is still manual manipulation of screenshots, the increase can be seen in video manipulation due to the growth of multimodal models. 

Mainstream social media platforms have a large impact on political accountability and fair elections. Recent elections have demonstrated that malicious actors can use social bots to subvert U.S. democracy through digital applications. Since LLM bots are capable of producing large quantities of credible text, they can be misused for disinformation purposes, specifically through automated campaigns with the goal of growing audience, and inflaming, or swaying opinion.

As Generative AI becomes more widely used, it is expected that investment in technological advancements by the private sector and the government should grow exponentially \cite{whiteHouse, goldmansachs}. In the absence of a wide-ranging regulatory framework synchronized with the development of applications and AI, many problems are arising. Considering how sophisticated LLMs are becoming, the differences between human-produced and AI-produced content have become extremely small. From grammar checking, news article generation, email drafting, and even website creation, AI is already being used in many areas of writing. However, sometimes AI-generated content can look like an attempt to imitate a human-like tone. Readers may feel uneasy about something that seems familiar and yet seems off. That phenomenon is called the Uncanny Valley. Masahiro Mori, a Japanese engineering professor, first proposed this hypothesis in 1970, saying that when a robot becomes more human-like, people's reactions will shift from affinity to revulsion. Since the proposal of the hypothesis, several studies have recreated or visualized the effect, tested its validity, and used perceptual, cognitive, or other types of analysis to investigate. These studies explored the uncanny valley primarily in the context of physical robots, digital avatars, and video chatbots that closely resemble humans. However, it is important to understand if and how LLM-generated text output can be used to manipulate and influence human behavior. 

57\% of people across of 19 countries around the world believe that social media is a good thing for democracy, however, 84\% of them also say that technological connectivity has made people easier to manipulate with false information and rumors \cite{pew1}. This manipulation can be done through the use of different technologies for the development of social bots. These bots can affect political discussion networks in several significant ways to amplify their messages. Pew Research Center found that most Americans are aware of social bots in a survey they conducted in 2018. However, only half of the respondents were at least "somewhat confident" that they could identify them, with only 7\% being "very confident". If those self-assessments are accurate, many users might already follow bots and share their content. Social media platforms often allow users to interact with one another and decide how to perceive their personalities and nature, yet in some cases, people may prefer to observe the conversations rather than engage in them. 

While previous research found that only 42\% of the time users are able to successfully identify the true nature of users on social media while interacting with them, we aim to understand if there is a difference between interactive and static identification of LLM-generated text content, as well as to understand the effects that LLM-generated text can have on humans. In light of recent developments, several key questions have emerged:
\begin{itemize}
    \item[] \textbf{RQ1: } How successful are humans in identifying the true nature of other users on social media (without the ability to interact with them)?

    \item[] \textbf{RQ2: } Are there patterns in how humans identify LLM-generated text content in social media discourse?

    \item[] \textbf{RQ3: } Is the Uncanny Valley observable in text as relates to the perception and identification of bots participating in social media discourse?
\end{itemize}

To address these questions, we performed a study with 1,095 participants. The survey was based on a dataset gathered by Radivojevic et al. \cite{radivojevic2024llms} during their experiment in which human and bot participants communicated on a social media platform without knowing the ratio or nature of participants. Our survey consisted of two randomly selected social media threads which included an initial post created by admin and the first five responses created by either human or bot participants in the previous experiment. Participants were tasked to select all responses that they believed were created by bots. While in the last decade, researchers have started to investigate how brain activity is related to the uncanny valley, the most common approach to studying the empirical basis of the phenomenon relies on self-reported subjective measurements \cite{diel2021meta, ratajczyk2022shape, katsyri2015review, wang2015uncanny}. To explore this effect in LLM-generated text, we then asked an additional question regarding the comfort level of participants while reading the human, bot, or both responses, based on their selections. Finally, we asked participants to provide a few reasons they believed some of the accounts were bots. 

Our findings can be summarized as the following contributions:

\begin{itemize}
    \item Despite foreknowledge of the presence of bots and humans and without the ability to interact with other users on the social media platform, humans are bad at identifying the true nature of posts. Participants were able to successfully identify the true nature of posts in the survey only 42\% of the time. 

    \item There are patterns in how participants perceive bot posts that can be classified into two groups: (1) human emotions, perception, and interaction and (2) evolution and mechanics of language and communication.

    \item Discomfort experienced when reading what participants believed were bot-generated posts align with the Uncanny Valley
    phenomenon, meaning that the participants found what they believed was bot-generated content less comfortable. The analysis also indicates that the participants’ perception aligns with their success rate and that the Uncanny Valley effect can be visible in both cases. 

    \item Additionally, we found that humans experience the Uncanny Valley effect when reading text generated by sophisticated LLMs. Despite our finding that humans are not able to identify bot-generated content, the sophistication of LLMs has still not reached a significant level of manipulating users completely and successfully on social media platforms.
    
\end{itemize}

Our work makes contributions to the community and the field of social and crowd computing, empirical investigations, as well as ethics and policy implications, focusing on the role of LLM-generated social bots in collective decisions and their effects on human perception when present on social media platforms. Our results are essential findings for the future of human-computer digital discourse and collaboration on social media and team platforms.  As the sophistication and complexity of bots grow alongside greater public access and ease of use; it becomes critical that we have both technical guardrails and policy guidance for the safe and effective growth of collective dialogue and decision-making with humans and bots.

\section{Related Work}

In this section, we synthesize literature on social bots, the Uncanny Valley, and their potential impact on digital discourse.

\subsection{Bots on Social Media}

Malicious bots have existed online since the rise of the internet. Studies show that around 50\% of all internet traffic comes from bot activity. Additionally, bad bots account for about 33\% of all internet traffic \cite{imperva2024}. Furthermore, human perception of bots remains largely negative with around 2/3rds of Americans having heard of bots, and 80\% believing they are used maliciously \cite{stocking2018social}.

Social bots have taken a spotlight within social media research due to their ability to influence public thinking by pushing specific agendas, particularly in a political setting. Social bots have already hosted successful disinformation campaigns in major political events such as United States presidential elections \cite{bessi2016social, badawy2018analyzing, shao2017spread} and the Brexit Referendum \cite{bastos2019brexit, howard2016bots}. Additionally, social bots use increased dramatically throughout the context of COVID-19 and played a pivotal role in the content users consumed about the pandemic \cite{himelein2021bots, suarez2022assessing, xu2022characterizing}. A study by \citet{seering2018social} finds that even when outnumbered on platforms such as Twitch, bots sent significantly more messages than humans. They note that at scale, these bots can easily influence user perception. A platform called Botivist developed by \citet{savage2016botivist} shows the strength of using Twitter social bots to engage users in discussions about various social issues and call users to action. These results highlight the risk of misuse, where similar platforms could be exploited to manipulate public opinion or spread misinformation. Such recent events and findings have elevated the urgency to develop anti-bot software or reliable detection systems to combat the rampant spread of misinformation.

Social bots serve primarily as actors seeking to push specific agendas. Furthermore, malicious agents often seek to spread biased information or misinformation that fits the narrative they support. No matter the function, social media bots operate under the direction of a botmaster who manages, curates, and programs all of the bot's activities \cite{orabi2020detection, stringhini2014harvester}. Social botnets refer to a group of social bots under the control of a single botmaster. These social botnets often aim to maintain or increase the amount of digital influence one has over a specific field, community, or platform \cite{zhang2016rise}. This is often performed by mimicking human behaviors to avoid detection and misdirecting users' attention away from relevant information \cite{abokhodair2015dissecting}. A case study performed by \citet{yang2023anatomy} demonstrates how a social botnet amplifies itself by reciprocal interaction such as retweets and likes. This social botnet was identified only by the accidental posting of self-revealing tweets generated by LLMs.

\subsection{Identifying Bots on Social Media}

Unlike social bots which mimic human behavior to influence interactions and discussions, traditional spambots are specifically designed to send unsolicited messages for advertising or phishing purposes. One of the most popular techniques used to differentiate bots from humans is the Completely Automated Public Turing Test to tell Humans and Computers Apart (CAPTCHA). This method however is considered outdated and can easily be defeated using human actors completing the CAPTCHA for bots or other algorithmic methods such as image recognition \cite{mori2003recognizing, bursztein2011text, chen2017survey}. Another technique used to detect spambots online is the HoneySpam application. This procedure baits spambots into attacking a specific system aimed at studying their behaviors and profiles \cite{hayati2009honeyspam, andreolini2005honeyspam}. Furthermore, some recent methods have been developed by \citet{ali2019detect} to detect traditional spambots via models based on graph convolutional neural networks.

One method of identifying social bots on various platforms employs the use of feature extraction. After identifying preliminary accounts deemed to be bots, the subset of features is fed into a shallow machine-learning model and then used to identify other accounts run by bots \cite{ilias2021detecting, ouni2022bots}. Other applications such as Botometer developed by \citet{yang2022botometer} employ the use of supervised learning to train a model that can predict with some certainty the possibility of various accounts being bots or not. Botometer analyzes the user's past 200 tweets along with other tweets that mention the bot account to determine their rating. However, this means that scores calculated by the Botometer are susceptible to change, especially for extremely active accounts \cite{rauchfleisch2020false}. Researchers have also recognized the potential of bot detection via deep learning models given their scalability and ability to keep pace with the rapid evolution of bots \cite{hayawi2023social, kudugunta2018deep, mcdermott2018botnet}. Some approaches such as the one suggested by \citet{ferreira2019uncovering} aim to teach media literacy to increase human recognition of bot accounts based on specific features such as grammatical content. The study states that educating users to recognize bots correctly will maintain a healthy social media environment even with social bots \cite{valtonen2019media}.

The rise of Large Language Models, such as ChatGPT, has made the proliferation of non-human-generated content much more viable on social media platforms. The synthesis of harmful content through LLMs' intentional training biases and specific prompt engineering enables the malicious use of human-like content \cite{zhuo2023red, barman2024dark}. Additionally, specific and appropriate prompts also enable social bots to interact with other accounts automatically rendering traditional methods of identification obsolete \cite{de2023chatgpt, ferrara2023social}. 

Currently, there exist two main methods of detecting LLM-generated text: black-box detection and white-box detection. Black-box detection methods rely purely on text samples from LLM generated and human text to train a model to differentiate the two. This detection method only requires a basic API level of access to the LLM. However, black-box models become less effective as models continue to evolve and become more sophisticated with each iteration \cite{tang2024science}. Furthermore, potential errors and weaknesses have been found when identifying various social bots using human perception and judgment and models trained on such premises. The "ground-truth label problem" hypothesized by \citet{kolomeets2024experimental} highlights how humans are unable to consistently identify social bots online, and systems trained on such human decisions also fail due to inheriting the same errors. Therefore, the overall effectiveness of systems trained on human labeling drops significantly compared to models trained purely on model validation. Experiments conducted by \citet{radivojevic2024llms} confirmed these hypotheses. Human users were not able to correctly identify and label LLM-based social bots when actively participating in discussions.

White-box detection consists of having complete access to the target LLM and imprinting hidden watermarks into the generated text. One major limitation however of white-box detection lies within its complexity. As the quality and effectiveness of the watermark increase, the integrity of the original text degrades \cite{jalil2009review, abdelnabi2021adversarial}. Additionally, malicious users could easily avoid this detection by using LLMs which are not watermarked.

\subsection{Investigating the Uncanny Valley}
The uncanny valley (UV) is described as the "proposed relation between the human likeness of an entity and the perceiver's affinity for it." As the entity's human likeness increases, the affinity also increases until a certain threshold. At this threshold, the affinity drastically dips until the entity begins to become indistinguishable from humans and begins to rise again. Traditionally, it has been studied mainly with respect to visual stimuli \cite{mori2012uncanny, wang2015uncanny}.

However, studies have indicated that UV extends to encompass human perception as a whole instead of limiting itself to just sight \cite{seibt2014sociable, gray2012feeling}. Multiple studies suggest specific discrepancies between various voices but fail to conclusively find a solid relationship between specific voice characteristics and UV \cite{do2022new, jansen2019discovering}. However, a recent study by \citet{diel2023vocal} suggests that the audio UV effect depends on the ``organicness" of the voice itself. Overall, the common consensus is that sound-based UV needs to be broken down and studied in a more categorical manner \cite{jansen2019discovering}.

Furthermore, studies also suggest a haptic UV exists in some capacity. An experiment run by Berger et al. found that incorporating the sense of touch with sight in virtual reality yields a similar effect to the standard UV. As the haptics continued to be included and specialized, the affinity of the user increased until a certain point \cite{berger2018uncanny}. Another study by \citet{d2019different} explored how the virtualization of sensory inputs affects self-attribution to avatars, revealing that virtualization generally decreases the sense of embodiment. The lowest sense of embodiment occurred when only one sensory input (either sight or touch) was virtual, causing revulsion and extending UV to avatar embodiment. The research emphasizes that matching the degree of virtualization for both visual and tactile stimuli is crucial for effective avatar representation.

Research findings involving the perception of algorithmic-generated content, however, remains divergent. Some studies show consumers are reluctant to trust algorithms for subjective tasks despite their superior performance. They find that increasing the human-like qualities of algorithms can enhance their acceptance and use \cite{castelo2019task}. Moreover, some assert that AI-generated or assisted writing inherently holds less value than content made solely by humans \cite{kulkarni2023llms}. Alternatively, other studies claim that an aversion to algorithms is instead a preference for human involvement \cite{morewedge2022preference}. An experiment done by \citet{zhang2023human} demonstrated that content generated by generative AI and augmented AI is often perceived as higher quality than that produced by human experts or augmented human teams. They claim this bias is driven by human favoritism towards human-produced content, which is partially mitigated by revealing the involvement of AI in the creation process. Research suggests, however, that understanding and transparency of how LLM and other AI generate their content will increase the amount of trust users have in such systems \cite{harper2024ml}. Regardless of algorithm aversion or human favoritism, a proportional relationship commonly appears in each experiment, which has parallels with the uncanny valley. 

As the human-like characteristics of algorithms and LLMs improve, so does the affinity the user has with it. Therefore, one must ask: "Can humans identify bots in social media discourse?", "Are there specific patterns and techniques they default to utilizing?" and "Does the uncanny valley exist and impact human perceptions in the context of LLM-generated text?". 

\begin{figure*}[t]
\centering
\includegraphics[width=0.95\textwidth]{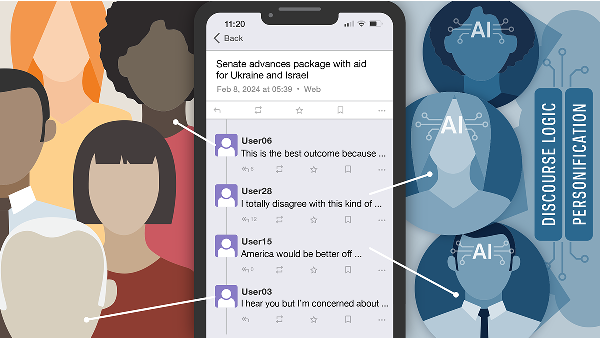}
\caption{Illustration of experimental framework where personified LLM bots participated in social discourse with humans. The resulting dataset which was made publicly available by \citet{radivojevic2024llms} was used to create a survey for this study.}
\label{fig1}
\end{figure*}

\section{Methodology}

We conducted a study on 1,095 participants to study the identification and perception of LLM bots in social media environments. We used Qualtrics to design the survey and Prolific to recruit participants. The participants were paid twelve USD per hour and took a median of about 6 minutes to complete the survey. We collected each participant's prolific id, consent form, time taken to complete the survey, age, sex, simplified ethnicity, and country of birth. Where available, we also collected their student and employment status. The participant summary is shown in Table \ref{tab:demographics}.

\begin{table}[h!]
\centering
\small
\caption{Demographic Data}
\label{tab:demographics}
\begin{tabular}{ |p{2cm}||p{1.5cm}p{1.5cm}p{1.5cm}p{1.5cm}p{1.5cm}p{1.5cm}|  }
 \hline
 Sex& \begin{tabular}{c} Female \\ (559) \end{tabular} & \begin{tabular}{c} Male \\ (536) \end{tabular} &  &  &  &  \\
 \hline
 Age& \begin{tabular}{c} 18-24 \\ (124) \end{tabular} & \begin{tabular}{c} 25-34 \\ (192) \end{tabular} & \begin{tabular}{c} 35-44 \\ (187) \end{tabular} & \begin{tabular}{c} 45-54 \\ (173) \end{tabular} & \begin{tabular}{c} >55 \\ (414) \end{tabular} & \begin{tabular}{c} NR \\ (5) \end{tabular}\\
 \hline
 Ethnicity& \begin{tabular}{c} White \\ (698) \end{tabular} & \begin{tabular}{c} Black \\ (131) \end{tabular} & \begin{tabular}{c} Mixed \\ (120) \end{tabular} & \begin{tabular}{c} Asian \\ (64) \end{tabular} & \begin{tabular}{c} Other \\ (82) \end{tabular} &  \\
 \hline
 Student& \begin{tabular}{c} Yes \\ (126) \end{tabular} & \begin{tabular}{c} No \\ (751) \end{tabular} & \begin{tabular}{c} NR \\ (218) \end{tabular} &  &  &  \\
 \hline
 Country& \begin{tabular}{c} U.S. \\ (1,013) \end{tabular} & \begin{tabular}{c} Other \\ (82) \end{tabular} &  &  &  &  \\
 \hline
 Employment& \begin{tabular}{c} Full-Time \\ (445) \end{tabular} & \begin{tabular}{c} Part-Time \\ (161) \end{tabular} & \begin{tabular}{c} Not-Paid \\ (146) \end{tabular} & \begin{tabular}{c} Unemployed \\ (92) \end{tabular} & \begin{tabular}{c} Other \\ (47) \end{tabular} & \begin{tabular}{c} NR \\ (204) \end{tabular} \\
 \hline
\end{tabular}
\end{table}

\subsection{Dataset Used for Survey Creation}

The survey was created using the dataset provided by Radivojevic et al \cite{radivojevic2024llms}. During the prior study, researchers conducted a real-time digital discourse experiment to study the impact of bots on social media. The research generated 24 discourses/threads gathered from three rounds of the experiment. They created 30 bot participants based on 10 personas gathered from literature on bots in global political discourse. Personas were developed and constructed on three different LLMs: GPT-4, Claude 2, and Llama 2 Chat by using prompt engineering techniques, resulting in 30 different bot accounts. 36 human participants were asked to interact with other users, both human and bots, on the platform. They were assigned a persona written in the same manner as the prompt that was used for bot construction and were tasked to engage with other participants' replies to foster a collaborative and interactive environment. The summary architecture used to create the dataset used for our study can be seen in Figure \ref{fig1}. The dataset consisted of 3,025 individual responses, of which 459 were human responses and 2,566 were bot responses.

\subsection{Study Design}

We used the Qualtrics online survey tool to design our study. For each of the 24 discourses, we ordered posts by timestamp (creation time) and selected the first five that were generated by human or bot participants after each initial topical post. Each participant was given two sets of social media posts with human, bot, or a mixture of responses. They were first tasked to click on the post for each response they believed to be from a bot. The ones that were not selected were considered human responses. Each of the two sets was followed by a Likert scale question, where a participant was asked to describe their level of comfort when reading posts coming from either human, bot, or both, depending on their selections in the previous post. As specific examples, if a participant believed all responses were human and did not select any bot replies, the bot comfort slider question did not appear. If a participant believed all responses were bots and selected all replies, the human comfort slider question did not appear. If a participant believed some responses were bot and some were humans, both comfort slider questions appeared. Finally, at the end of the survey participants were asked to explain what characteristics of posts gave them the impression or perception that they were bot posts.

\section{Analysis and Results}

In this section, we provide an analysis of results in response to our research questions. First, we investigate how well users of a statically presented social media discourse can identify bot posts (RQ1). Then, we identify patterns in how humans perceive and identify LLM-generated text (RQ2). Finally, we explore if the Uncanny Valley effect applies to text conversations (RQ3). 

\subsection{How successful are humans in identifying the true nature of other users on social media
without the ability to interact with them? (RQ1)}

\begin{figure*}[t]
\centering
\includegraphics[width=0.6\textwidth]{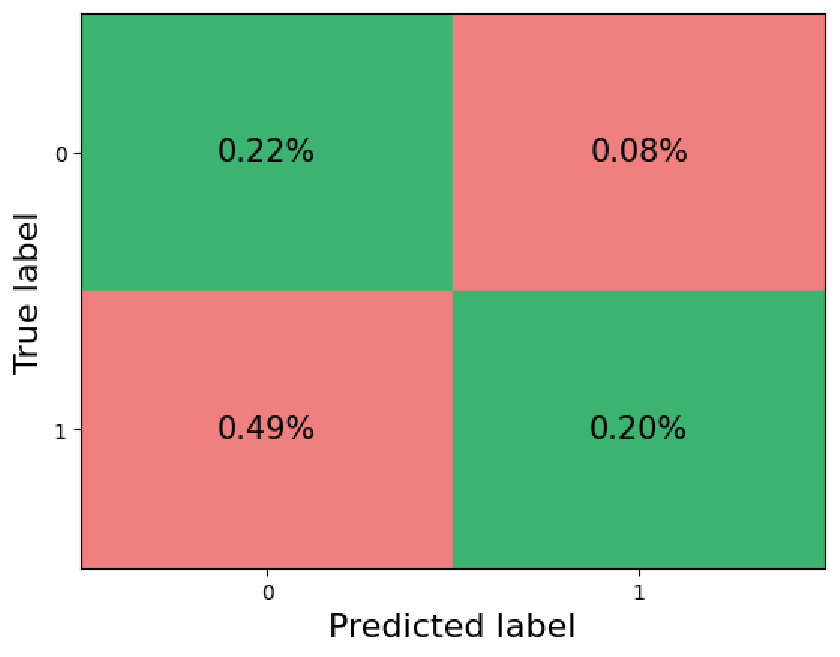}
\caption{Confusion matrix of predicted and actual nature of posts predicted by participants. 0 = human, 1 = bot. The results indicate that participants are successful only 42\% of the time when tasked to identify the true nature of posts despite foreknowledge of the presence of bot and human posts in the survey.}
\label{fig2}
\end{figure*}

Participants were asked to select all posts they believed were written by bots. To calculate the overall performance of participants in identifying bot posts, we compare the actual nature of options in the survey with those predicted by participants. The confusion matrix in Figure \ref{fig2} represents the results of the participant's performance in identifying the true nature of posts in the survey. The results indicate 42\% accuracy in identifying the true nature, with a high false negative rate of 49\% indicating participants incorrectly identified bots as humans. The previous experiment conducted by \citet{radivojevic2024llms} showed that humans were only 42\% accurate when attempting to identify the true nature of participants with the ability to interact with them. While they focus on the nature of users on the platform, we focus on the nature of posts and the ability of content to influence and persuade survey participants without the ability to interact or gather more information about the posts they evaluate. The poor detection accuracy of 42\% in this experiment shows the dangers of LLM-created content that can be used by propagandists to manipulate users on social media. Our findings suggest that despite the foreknowledge of the presence of bot and human replies in the survey, humans are bad at identifying the true nature of replies with and without the ability to interact with other users on social media. The 42\% accuracy from two separate experiments was surprisingly close agreement. However, distributions for the confusion matrices differed. 

\subsection{Patterns in How Humans Identify LLM-generated Text Content (RQ2)}


\begin{figure*}[t]
\centering
\includegraphics[width=0.7\textwidth]{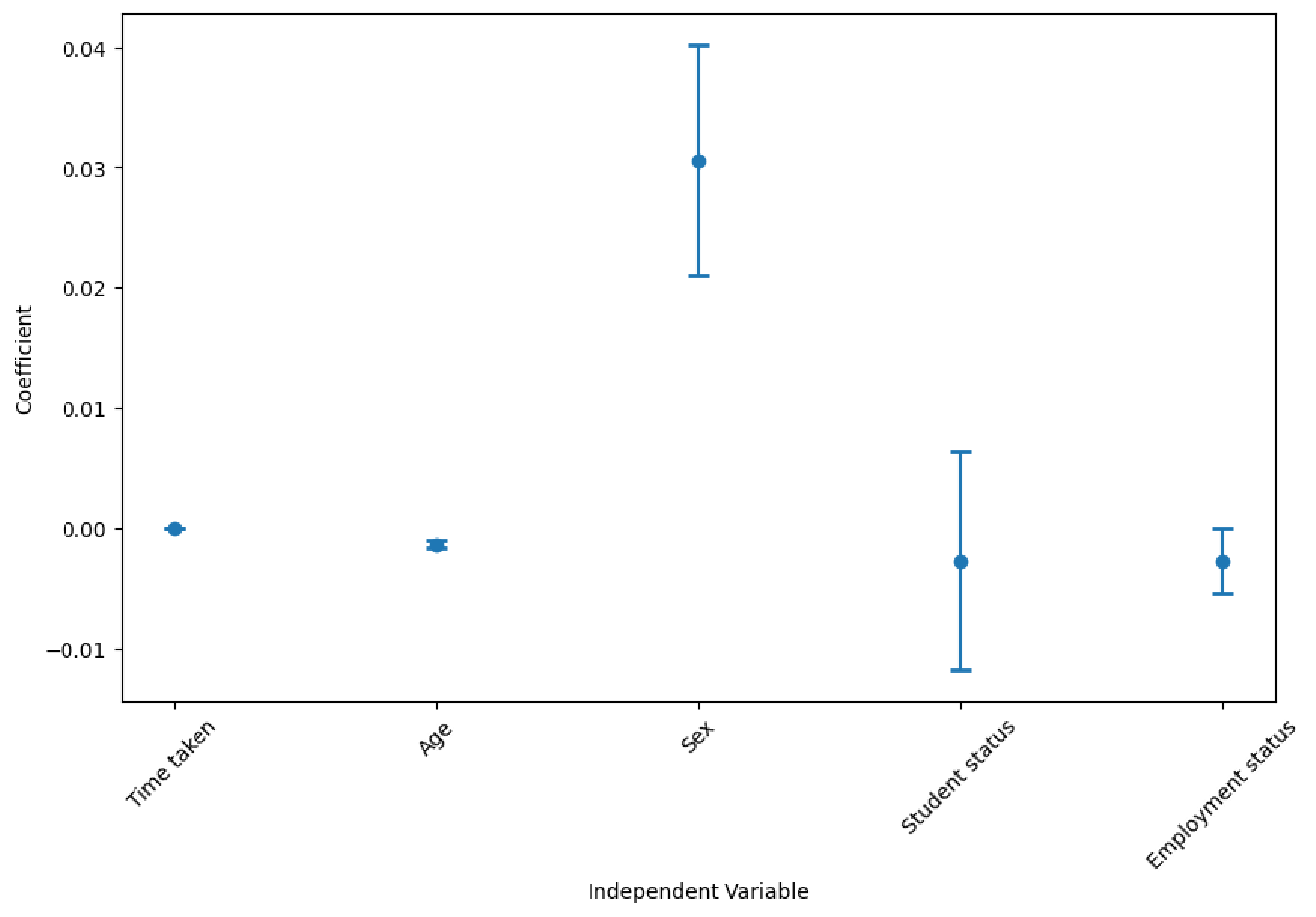}
\caption{Coefficient plot showing the results of linear regression and the relationship between independent variables and success rate as the dependent variable.}
\label{fig4}
\end{figure*}

To identify the relationships between demographic variables in the experiment and success rate we performed exploratory data analysis and calculated the correlation matrix. We considered variables such as prediction, success rate (which is defined as the success in predicting the true nature of replies in the survey) and demographics of participants (age, sex, simplified ethnicity, student status, employment status). 

\subsubsection{Success Rate as the Dependent Variable}

We performed an OLS analysis to investigate the relationship between independent variables and success rate as the dependent variable. The results are shown on the coefficient plot in Figure \ref{fig4}. We also performed a Chi-square test for independence. Our findings with a Chi-square statistic of 12.53 and a $p$-value of 0.0003 indicate that the success rate has a small dependence on sex in our data, with male participants being successful 44.4\% of the time while female participants were successful 41.1\% of the time. 

We performed the same analysis to calculate the success rate based on the age groups of participants in the survey. Results indicate no significant difference among age groups in identifying the true nature of replies. The results indicate 42.3\% success rate for the age group 18-24, 47.2\% success rate for the age group 25-34, 42.9\% success rate for the age group 35-44, 44.3\% success rate for the age group 45-54, 39.8\% success rate for participants older than 55.

We then performed logistic regression analysis to explore relationships between participant's success rate and their demographics. The results indicate that age and sex are significant predictors of participants' success in the identification process, while other variables are not. Results are shown in Table \ref{tab:logits2}.

\begin{table}[t]
\begin{center}
\caption{Logistic regression analysis of whether participants are more successful in selecting human or bot.}
\vspace{-.1cm}
\small
\label{tab:logits2}
\begin{tabular}{rrrrrl}
\hline
& \textbf{$\beta$} & \textbf{Std. Err} & \textbf{t-value} & \textbf{Pr($>$$|$t$|$)} &   \\ 
\hline
Const. & -0.1017 & 0.299 & -0.340 & 0.734 &  \\ 
Time taken & 4.576e-05 & 6.97e-05 & 0.657 & 0.511 &  \\ 
Age & -0.0057 & 0.001 & -4.511 & 0.000 & *** \\ 
Sex & 0.1335 & 0.039 & 3.439 & 0.001 & *** \\ 
Ethnicity & 0.0229 & 0.016 & 1.460 & 0.144 &  \\ 
Student status & -0.0077 & 0.037 & -0.206 & 0.836 &  \\ 
Employment status & -0.0114 & 0.011 & -1.016 & 0.309 &  \\ 
    \hline
\multicolumn{6}{r}{*$p$<0.05; **$p$<0.01; ***$p$<0.001 } \\
\end{tabular}
\end{center}
\end{table}

\subsubsection{Prediction as the Dependent Variable}

To explore the relationship between participant's prediction and their demographics we performed logistic regression analysis. The results indicate that age, sex, and simplified ethnicity are significant predictors of the dependent variable, in this case, user prediction if a reply is human or bot-created, with a $p$ < 0.05, while the time taken to complete the survey, country of birth, and nationality are not. Significant intercept suggest that the baseline log-odds is not zero. Results of the logistic regression are shown in Table \ref{tab:logits1}.

\begin{table}[t]
\begin{center}
\caption{Logistic regression analysis of whether participants tend to predict human or bot.}
\vspace{-.1cm}
\small
\label{tab:logits1}
\begin{tabular}{rrrrrl}
\hline
& \textbf{$\beta$} & \textbf{Std. Err} & \textbf{t-value} & \textbf{Pr($>$$|$t$|$)} &   \\ 
\hline
Const. & -1.0353 & 0.335 & -3.090 & 0.002 & ** \\ 
Time taken & 0.0001 & 7.61e-05 & 1.907 & 0.057 &  \\ 
Age & -0.0141 & 0.001 & -10.000 & 0.000 & *** \\ 
Sex & 0.2429 & 0.043 & 5.677 & 0.000 & *** \\ 
Ethnicity & 0.0530 & 0.017 & 3.039 & 0.002 & ** \\ 
Student status & -0.0227 & 0.041 & -0.561 & 0.575 &  \\ 
Employment status & -0.0174 & 0.012 & -1.405 & 0.160 &  \\ 
    \hline
\multicolumn{6}{r}{*$p$<0.05; **$p$<0.01; ***$p$<0.001 } \\
\end{tabular}
\end{center}
\end{table}

\begin{figure*}[t]
\centering
\includegraphics[width=0.7\textwidth]{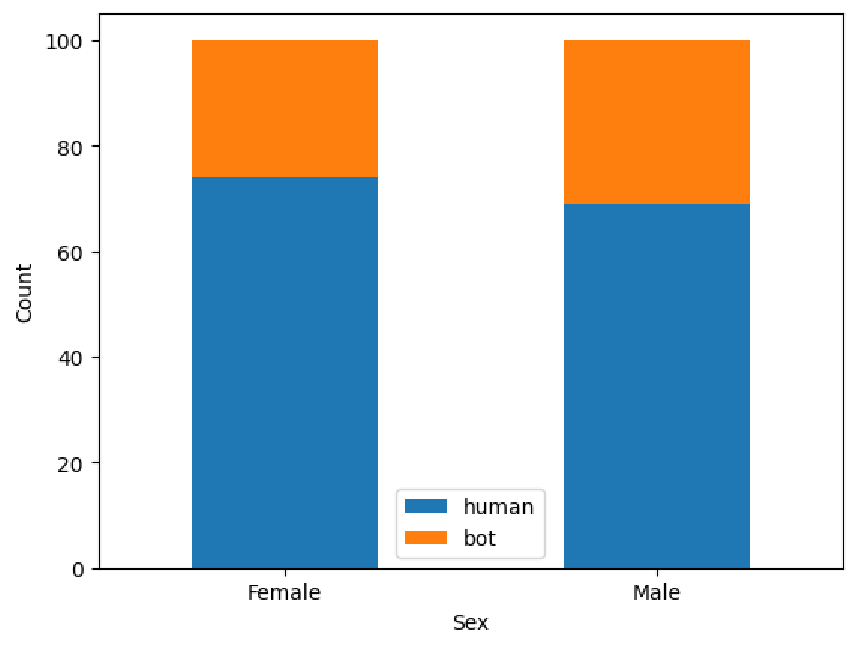}
\caption{Tendency of male and female participants to predict posts as bot or human.}
\label{fig5}
\end{figure*}

We then calculated the tendency of male and female participants to select replies as bot or humans. The results indicate that female participants tend to identify posts in the survey as human replies more often than male participants, while male participants tend to identify posts as bot participants more often than female participants. Results are shown in Figure \ref{fig5}. Both groups of participants in the survey tend to identify replies as human-created content. 

\subsubsection{Analysis of bot indicators based on participant responses.}

To explore and identify patterns in what participants believed was bot-generated content, we performed an in-depth analysis of textual responses provided by participants who were asked to explain what are some indicators that led them to select certain replies as bot-generated. We first applied the Latent Dirichlet Allocation (LDA) model to identify topics and group words based on their weights into a predefined number of clusters. We utilized the LDA model as it combines an inductive approach with quantitative computations of large-size textual data. Before running the LDA model, the text from 1,095 participant responses was pre-processed (stop and infrequent words were removed, lemmatizing, tokenizing, etc.) and a document matrix was created. To determine the initial number of topics we referred to the previous work by \citet{radivojevic2024llms} and responses provided in the initial experiment. Based on qualitative analysis, we determined the initial number of topics \emph{(K)} the LDA model should classify to be four. The initial results yielded 10 words for each topic. We then prompted GPT-4o LLM to define and generate the topic titles based on the words identified by LDA. The Prompt looked as follows:\\

\emph{You are an advanced AI specializing in natural language processing. 
Your task is to analyze a list of identified words after Latent Dirichlet Allocation analysis and define specific topics for each word group. 
Here is a list of identified most common words from user comments: 
Topic 1: felt, language, human, emojis, used, use, old, formal, words, year
Topic 2: human, really, sure, could, think, trying, real, last, first, wrong
Topic 3: human, think, user, emotion, made, responses, response, point, believe, sound
Topic 4: emojis, use, human, way, think, hashtags, people, feel, used, many
Steps to follow:
Identify the four main topics discussed based on the most frequent word described here.}\\

We then prompted the same model to analyze the topics identified and suggest a consolidated list of fewer topics that could streamline the analysis, as some of the themes overlapped. The model then yielded a list of three topics as follows:

\emph{**Evolution of Language and Communication Tools**
   \underline{Words}: felt, language, emojis, used, use, old, formal, words, year, hashtags
   \underline{Description}: This topic encompasses how language and communication have transformed over time, including the adoption of digital communication tools like emojis and hashtags, alongside the gradual shift from formal to more informal methods of expressing oneself.}
   
\emph{**Human Nature and Perception**
   \underline{Words}: human, really, sure, could, think, trying, real, last, first, wrong
   - \underline{Description}: This topic delves into the intricacies of human nature, exploring how people perceive reality, their thought processes, and their efforts to distinguish between right and wrong.}
   
\emph{**User Interaction and Emotional Responses**
   \underline{Words}: human, think, user, emotion, made, responses, response, point, believe, sound, feel, people, use, way 
   \underline{Description}: This topic focuses on how users interact with one another and respond emotionally in various contexts. It also covers how these interactions sound authentic or manufactured, and the role people's feelings play}

We then calculated the coherence score for LDA. Due to the moderate score of 0.45, we then performed an additional analysis and identified the 100 most frequent words in the dataset produced by participants in the survey. We then prompted GPT-4o to define 3 topics based on the list of most frequent words. The output looked as follows:

\emph{**Human Emotions and Perceptions:**
   \underline{Words}: human, felt, feel, real, personal, person, emotion, response, emotional, emotions, feelings, feeling, sense.
   \underline{Description}: This topic involves conversations about human emotions, feelings, and perceptions. It encompasses how people express and interpret emotions in writing and in interactions, focusing on the subjective experience of being human.}

\emph{**Language and Communication:**
   \underline{Words:} think, use, used, written, responses, trying, words, language, grammar, sentences, sentence, tone, comments, wording, punctuation, say, said, sound, sounds, sounded, wording.
   \underline{Description} This topic centers around the mechanics of language and communication. It covers the use of language, grammar, sentence structure, tone, and punctuation in conveying ideas and emotions effectively.}

\emph{**Artificial Intelligence and Its Interaction with Humans:**
  \underline{Words:} AI, humans, user, generated, natural, unnatural, overly, generic, believe, makes, might, see, based, similar, likely, structure, using, doesn't, facts
    \underline{Description:}  This topic explores the intersection between artificial intelligence (AI) and human users. Discussions include how AI-generated content compares to human-created content in terms of naturalness and authenticity. It addresses the effectiveness, limitations, and perceptions of AI, including whether AI responses feel generic or overly formal. Additionally, it considers the trustworthiness of AI-generated information and how well AI can mimic human language and behavior. This topic also dives into the structural and factual accuracy of AI outputs and how users interact with and respond to AI systems.}

Finally, we prompted the model to analyze, combine, and produce the final topics based on the information mentioned above. The final two topics identified in the dataset are as follows:

\begin{itemize}
\item \textbf{Human Emotions, Perception, and Interaction:}

   **Keywords:** human, felt, real, personal, person, emotion, response, emotional, emotions, feelings, feeling, sense, really, sure, think, trying, last, first, wrong, believe, point, way
   
   **Description:** This topic delves into how people perceive reality, experience and express emotions, and interact with each other. It explores the complexities of human nature, including thought processes and efforts to distinguish right from wrong. It also covers how people’s feelings and emotions shape their responses and interactions.

\item \textbf{Evolution and Mechanics of Language and Communication:}

   **Keywords:** language, use, used, written, words, grammar, sentences, sentence, tone, punctuation, emojis, year, hashtags, old, formal, comments, wording, say, said, sound, sounds, sounded

    **Description:** This topic captures how language and communication have evolved over time, including shifts from formal to informal expressions and the incorporation of digital tools like emojis and hashtags. It also addresses the mechanics of language, such as grammar, sentence structure, tone, and punctuation.
\end{itemize}
To include a human in the loop and confirm the topics identified by LDA and GPT-4o, we randomly selected a subset of 10\% of comments from the dataset produced by participants and performed a reflective thematic analysis to compare human perception of text versus GPT-4o perception of text and topics identified. The analysis indicates that 94\% of the time human evaluation aligned with the topics predicted by the model. 

\subsection{Is the Uncanny Valley observable in text as relates to the perception and identification of bots participating in social media discourse? (RQ3)}


Each of the two sets was followed by a Likert scale question, where a participant was asked to describe their level of comfort when reading posts coming from either human, bot, or both, depending on their selections in the previous post. The Likert scale question for both human and bot had the following options:
\begin{itemize}
    \item[1 - ] Very uncomfortable. You felt extremely uneasy, anxious, or distressed during the interaction.

    \item[2 - ]Uncomfortable. You felt somewhat uneasy or bothered during the interaction. The experience was unpleasant, but not to the extreme.

    \item[3 - ]Neutral. You felt neither comfortable nor uncomfortable during the interaction. The experience was neither pleasant nor unpleasant. You felt indifferent and had no strong feelings either way.

    \item[4 - ]Comfortable. You felt generally at ease and relaxed during the interaction. The experience was pleasant, and you were comfortable and without any significant concerns.

    \item[5 - ]Very comfortable. You felt extremely at ease, relaxed, and content during the interaction.
\end{itemize}

We performed an ordinal regression analysis to identify relationships between independent variables and the dependent variables related to the Likert scale. The survey consisted of two Likert scales that were programmed to appear based on the participants' selection. First, we considered a human slider for the analysis. Results in Table \ref{tab:ordinalHuman} indicate that there is a significant relationship between age, ethnicity, and student status and the dependent variable human slider, while the thresholds of the Likert scale provide insights into how the ordered categories of the dependent variable Human Likert scale are separated, suggesting a clear separation between the categories.

\begin{table}[t]
\begin{center}
\caption{Ordinal regression analysis showing the relationship between an independent variable and Human Likert scale as the dependent variable.}
\vspace{-.1cm}
\small
\label{tab:ordinalHuman}
\begin{tabular}{rrrrrl}
\hline
& \textbf{$\beta$} & \textbf{Std. Err} & \textbf{t-value} & \textbf{Pr($>$$|$t$|$)} &   \\ 
\hline
Success rate & 0.0037 & 0.020 & 0.184 & 0.854 &  \\ 
Age & 0.0026 & 0.001 & 3.960 & 0.000 & *** \\ 
Sex & -0.0226 & 0.020 & -1.115 & 0.265 &  \\ 
Ethnicity & 0.0668 & 0.021 & 3.170 & 0.002 & ** \\ 
Student status & -0.0715 & 0.019 & -3.709 & 0.000 & *** \\ 
Employment status & 0.0002 & 0.006 & 0.035 & 0.972 &  \\ 
0/1 & -1.9857 & 0.074 & -26.807 & 0.000 & *** \\
1/2 & -0.8137 & 0.050 & -16.301 & 0.000 & *** \\
2/3 & -0.4460 & 0.027 & -16..804 & 0.000 & *** \\
3/4 & -0.0279 & 0.014 & -1.946 & 0.000 & *** \\
4/5 & -0.1516 & 0.015 & -9.949 & 0.000 & *** \\
    \hline
\multicolumn{6}{r}{*$p$<0.05; **$p$<0.01; ***$p$<0.001 } \\
\end{tabular}
\end{center}
\end{table}

Then, we considered a bot slider as the dependent variable and the results indicate that there is a significant relationship between all predictor variables and the dependent variable bot slider. Once again, the threshold coefficient indicates that there is a clear separation between categories offered in the Likert scale. 

\begin{table}[t]
\begin{center}
\caption{Ordinal regression analysis showing the relationship between an independent variable and Bot Likert scale as the dependent variable.}
\vspace{-.1cm}
\small
\label{tab:ordinalbot}
\begin{tabular}{rrrrrl}
\hline
& \textbf{$\beta$} & \textbf{Std. Err} & \textbf{t-value} & \textbf{Pr($>$$|$t$|$)} &   \\ 
\hline
Success rate & 0.3011 & 0.021 & 14.500 & 0.000 & *** \\ 
Age & 0.1714 & 0.021 & 8.247 & 0.000 & *** \\ 
Sex & -0.0083 & 0.001 & -12.240 & 0.000 & *** \\ 
Ethnicity & 0.1060 & 0.022 & 4.883 & 0.000 & *** \\ 
Student status & -0.0603 & 0.020 & -3.067 & 0.002 & * \\ 
Employment status & -0.0123 & 0.006 & -2.060 & 0.039 & *  \\ 
0/1 & -0.3783 & 0.071 & -5.303 & 0.000 & *** \\
1/2 & -2.1054 & 0.045 & -47.190 & 0.000 & *** \\
2/3 & -1.0631 & 0.025 & -43.026 & 0.000 & *** \\
3/4 & -0.3131 & 0.017 & -18.710 & 0.000 & *** \\
4/5 & -0.5989 & 0.025 & -23.877 & 0.000 & *** \\
    \hline
\multicolumn{6}{r}{*$p$<0.05; **$p$<0.01; ***$p$<0.001 } \\
\end{tabular}
\end{center}
\end{table}

To explore the Uncanny Valley in text, we performed a $t$-test to compare the comfort levels when evaluating what participants believed to be bot and human-generated content. We first calculated the mean scores of the human Likert scale and bot Likert scale in the survey. Our findings indicate $t$-statistics as high as 70.2, suggesting a large difference in the comfort levels reported for bot versus human replies. We then report the $p$-value that indicates the probability that the observed differences in comfort levels could occur by random chance if there were no actual differences between the groups. The $p$-value of near 0 indicates that the difference in comfort levels is highly statistically significant. 

We also calculated if there is the Uncanny Valley effect when participants were successful in identifying the true nature of replies. The $t$-statistics as high as 36.7 suggest a large difference in the comfort level reported for bot versus human replies. The $p$-value of 0 indicates the statistical significance of this finding. 

Lastly, we calculated the number of participants who correctly identified all responses in one survey question out of two. A total of 54 participants out of 1,095 were successful in identifying the true nature of the posts in the survey. Only two participants were successful in identifying the true nature of all replies to both questions. The $t$-statistics of 3.1 with $p$=0.003 indicates a moderate but statistical difference in comfort levels reported for successfully identified bot and human posts. 

The overall findings regarding the Uncanny Valley indicate that the discomfort experienced when reading what participants believed were bot-generated posts aligns with the Uncanny Valley phenomenon, meaning that the participants find what they believed was bot-generated content less comfortable. The analysis also indicates that the participants' perception aligns with their success rate and that the Uncanny Valley effect can be visible in both cases. Our findings also suggest that 54 participants who were 100\% accurate in the identification of the true nature of posts also experienced the Uncanny Valley effect.

\section{Limitations and Future Work}

Several limitations in our study can inform future research. First, the data used for survey creation comes from the previous experiment in which the bots were prompt-engineered using GPT-4, Claude 2, and Llama 2 Chat models. Different models with different techniques used (e.g. fine-tuning) might yield different outcomes, which can potentially affect our findings. Next, a larger sample size might provide a more accurate representation of the population. Further, implementing different methods along with the self-report for the Uncanny Valley analysis, as well as performing the Uncanny Valley experiment in the dynamic social media environment with LLM-generated content can potentially confirm or reject our findings. Finally, our work does not provide qualitative human reflective analysis on the entire dataset, but rather on a 10\% subset.

\section{Conclusion}

A user's affinity with algorithms and LLMs increases as they develop human-like characteristics. Because of that, it is important to understand if and how LLM-generated output can be used to influence human behavior, especially when users on social media platforms do not have the ability to interact with specific content or users. To understand the human perception of LLM-generated content in the social media environment, we conducted a study that yields a concerning finding that humans are bad at identifying the true nature of posts. Our findings which indicate that there are patterns and predictors in how humans select and identify content can contribute to the development of educational tools that can be used in teaching people how to identify the attempt of manipulation in the digital environment. We also find that humans experience the Uncanny Valley effect in the text that is generated by sophisticated LLMs which, despite our finding that humans are bad at identifying bot-generated content, indicates that the level of LLM sophistication has still not reached a level to completely manipulate users on social media platforms. However, it is important to note that the data used in the survey was not a product of fine-tuned models but rather prompt-engineered ones, which can potentially implicate different findings regarding the Uncanny Valley effect. 

\section{Acknowledgements}

The authors would like to thank Dr. Tim Weninger for providing valuable feedback and suggestions that helped refine and improve this work. We thank the ND Service Desk and Martin Klubeck for their assistance in building the static Qualtrics survey. Lastly, we thank Susan Joy Nduta Gicheha for her help in data analytics.

\bibliographystyle{ACM-Reference-Format}
\bibliography{sample-base}


\begin{thebibliography}{61}


\ifx \showCODEN    \undefined \def \showCODEN     #1{\unskip}     \fi
\ifx \showDOI      \undefined \def \showDOI       #1{#1}\fi
\ifx \showISBNx    \undefined \def \showISBNx     #1{\unskip}     \fi
\ifx \showISBNxiii \undefined \def \showISBNxiii  #1{\unskip}     \fi
\ifx \showISSN     \undefined \def \showISSN      #1{\unskip}     \fi
\ifx \showLCCN     \undefined \def \showLCCN      #1{\unskip}     \fi
\ifx \shownote     \undefined \def \shownote      #1{#1}          \fi
\ifx \showarticletitle \undefined \def \showarticletitle #1{#1}   \fi
\ifx \showURL      \undefined \def \showURL       {\relax}        \fi
\providecommand\bibfield[2]{#2}
\providecommand\bibinfo[2]{#2}
\providecommand\natexlab[1]{#1}
\providecommand\showeprint[2][]{arXiv:#2}

\bibitem[Abdelnabi and Fritz(2021)]%
        {abdelnabi2021adversarial}
\bibfield{author}{\bibinfo{person}{Sahar Abdelnabi} {and} \bibinfo{person}{Mario Fritz}.} \bibinfo{year}{2021}\natexlab{}.
\newblock \showarticletitle{Adversarial watermarking transformer: Towards tracing text provenance with data hiding}. In \bibinfo{booktitle}{\emph{2021 IEEE Symposium on Security and Privacy (SP)}}. IEEE, \bibinfo{pages}{121--140}.
\newblock


\bibitem[Abokhodair et~al\mbox{.}(2015)]%
        {abokhodair2015dissecting}
\bibfield{author}{\bibinfo{person}{Norah Abokhodair}, \bibinfo{person}{Daisy Yoo}, {and} \bibinfo{person}{David~W McDonald}.} \bibinfo{year}{2015}\natexlab{}.
\newblock \showarticletitle{Dissecting a social botnet: Growth, content and influence in Twitter}. In \bibinfo{booktitle}{\emph{Proceedings of the 18th ACM conference on computer supported cooperative work \& social computing}}. \bibinfo{pages}{839--851}.
\newblock


\bibitem[Ali~Alhosseini et~al\mbox{.}(2019)]%
        {ali2019detect}
\bibfield{author}{\bibinfo{person}{Seyed Ali~Alhosseini}, \bibinfo{person}{Raad Bin~Tareaf}, \bibinfo{person}{Pejman Najafi}, {and} \bibinfo{person}{Christoph Meinel}.} \bibinfo{year}{2019}\natexlab{}.
\newblock \showarticletitle{Detect me if you can: Spam bot detection using inductive representation learning}. In \bibinfo{booktitle}{\emph{Companion proceedings of the 2019 world wide web conference}}. \bibinfo{pages}{148--153}.
\newblock


\bibitem[Andreolini et~al\mbox{.}(2005)]%
        {andreolini2005honeyspam}
\bibfield{author}{\bibinfo{person}{Mauro Andreolini}, \bibinfo{person}{Alessandro Bulgarelli}, \bibinfo{person}{Michele Colajanni}, {and} \bibinfo{person}{Francesca Mazzoni}.} \bibinfo{year}{2005}\natexlab{}.
\newblock \showarticletitle{HoneySpam: Honeypots Fighting Spam at the Source.}
\newblock \bibinfo{journal}{\emph{SRUTI}}  \bibinfo{volume}{5} (\bibinfo{year}{2005}), \bibinfo{pages}{11--11}.
\newblock


\bibitem[Badawy et~al\mbox{.}(2018)]%
        {badawy2018analyzing}
\bibfield{author}{\bibinfo{person}{Adam Badawy}, \bibinfo{person}{Emilio Ferrara}, {and} \bibinfo{person}{Kristina Lerman}.} \bibinfo{year}{2018}\natexlab{}.
\newblock \showarticletitle{Analyzing the digital traces of political manipulation: The 2016 Russian interference Twitter campaign}. In \bibinfo{booktitle}{\emph{2018 IEEE/ACM international conference on advances in social networks analysis and mining (ASONAM)}}. IEEE, \bibinfo{pages}{258--265}.
\newblock


\bibitem[Barman et~al\mbox{.}(2024)]%
        {barman2024dark}
\bibfield{author}{\bibinfo{person}{Dipto Barman}, \bibinfo{person}{Ziyi Guo}, {and} \bibinfo{person}{Owen Conlan}.} \bibinfo{year}{2024}\natexlab{}.
\newblock \showarticletitle{The dark side of language models: Exploring the potential of llms in multimedia disinformation generation and dissemination}.
\newblock \bibinfo{journal}{\emph{Machine Learning with Applications}} (\bibinfo{year}{2024}), \bibinfo{pages}{100545}.
\newblock


\bibitem[Bastos and Mercea(2019)]%
        {bastos2019brexit}
\bibfield{author}{\bibinfo{person}{Marco~T Bastos} {and} \bibinfo{person}{Dan Mercea}.} \bibinfo{year}{2019}\natexlab{}.
\newblock \showarticletitle{The Brexit botnet and user-generated hyperpartisan news}.
\newblock \bibinfo{journal}{\emph{Social science computer review}} \bibinfo{volume}{37}, \bibinfo{number}{1} (\bibinfo{year}{2019}), \bibinfo{pages}{38--54}.
\newblock


\bibitem[Berger et~al\mbox{.}(2018)]%
        {berger2018uncanny}
\bibfield{author}{\bibinfo{person}{Christopher~C Berger}, \bibinfo{person}{Mar Gonzalez-Franco}, \bibinfo{person}{Eyal Ofek}, {and} \bibinfo{person}{Ken Hinckley}.} \bibinfo{year}{2018}\natexlab{}.
\newblock \showarticletitle{The uncanny valley of haptics}.
\newblock \bibinfo{journal}{\emph{Science Robotics}} \bibinfo{volume}{3}, \bibinfo{number}{17} (\bibinfo{year}{2018}), \bibinfo{pages}{eaar7010}.
\newblock


\bibitem[Bessi and Ferrara(2016)]%
        {bessi2016social}
\bibfield{author}{\bibinfo{person}{Alessandro Bessi} {and} \bibinfo{person}{Emilio Ferrara}.} \bibinfo{year}{2016}\natexlab{}.
\newblock \showarticletitle{Social bots distort the 2016 US Presidential election online discussion}.
\newblock \bibinfo{journal}{\emph{First monday}} \bibinfo{volume}{21}, \bibinfo{number}{11-7} (\bibinfo{year}{2016}).
\newblock


\bibitem[Bursztein et~al\mbox{.}(2011)]%
        {bursztein2011text}
\bibfield{author}{\bibinfo{person}{Elie Bursztein}, \bibinfo{person}{Matthieu Martin}, {and} \bibinfo{person}{John Mitchell}.} \bibinfo{year}{2011}\natexlab{}.
\newblock \showarticletitle{Text-based CAPTCHA strengths and weaknesses}. In \bibinfo{booktitle}{\emph{Proceedings of the 18th ACM conference on Computer and communications security}}. \bibinfo{pages}{125--138}.
\newblock


\bibitem[Castelo et~al\mbox{.}(2019)]%
        {castelo2019task}
\bibfield{author}{\bibinfo{person}{Noah Castelo}, \bibinfo{person}{Maarten~W Bos}, {and} \bibinfo{person}{Donald~R Lehmann}.} \bibinfo{year}{2019}\natexlab{}.
\newblock \showarticletitle{Task-dependent algorithm aversion}.
\newblock \bibinfo{journal}{\emph{Journal of Marketing Research}} \bibinfo{volume}{56}, \bibinfo{number}{5} (\bibinfo{year}{2019}), \bibinfo{pages}{809--825}.
\newblock


\bibitem[Chen et~al\mbox{.}(2017)]%
        {chen2017survey}
\bibfield{author}{\bibinfo{person}{Jun Chen}, \bibinfo{person}{Xiangyang Luo}, \bibinfo{person}{Yanqing Guo}, \bibinfo{person}{Yi Zhang}, {and} \bibinfo{person}{Daofu Gong}.} \bibinfo{year}{2017}\natexlab{}.
\newblock \showarticletitle{A Survey on Breaking Technique of Text-Based CAPTCHA}.
\newblock \bibinfo{journal}{\emph{Security and communication networks}} \bibinfo{volume}{2017}, \bibinfo{number}{1} (\bibinfo{year}{2017}), \bibinfo{pages}{6898617}.
\newblock


\bibitem[De~Angelis et~al\mbox{.}(2023)]%
        {de2023chatgpt}
\bibfield{author}{\bibinfo{person}{Luigi De~Angelis}, \bibinfo{person}{Francesco Baglivo}, \bibinfo{person}{Guglielmo Arzilli}, \bibinfo{person}{Gaetano~Pierpaolo Privitera}, \bibinfo{person}{Paolo Ferragina}, \bibinfo{person}{Alberto~Eugenio Tozzi}, {and} \bibinfo{person}{Caterina Rizzo}.} \bibinfo{year}{2023}\natexlab{}.
\newblock \showarticletitle{ChatGPT and the rise of large language models: the new AI-driven infodemic threat in public health}.
\newblock \bibinfo{journal}{\emph{Frontiers in public health}}  \bibinfo{volume}{11} (\bibinfo{year}{2023}), \bibinfo{pages}{1166120}.
\newblock


\bibitem[Diel and Lewis(2023)]%
        {diel2023vocal}
\bibfield{author}{\bibinfo{person}{Alexander Diel} {and} \bibinfo{person}{Michael Lewis}.} \bibinfo{year}{2023}\natexlab{}.
\newblock \showarticletitle{The vocal uncanny valley: Deviation from typical organic voices best explains uncanniness.}
\newblock  (\bibinfo{year}{2023}).
\newblock


\bibitem[Diel et~al\mbox{.}(2021)]%
        {diel2021meta}
\bibfield{author}{\bibinfo{person}{Alexander Diel}, \bibinfo{person}{Sarah Weigelt}, {and} \bibinfo{person}{Karl~F Macdorman}.} \bibinfo{year}{2021}\natexlab{}.
\newblock \showarticletitle{A meta-analysis of the uncanny valley's independent and dependent variables}.
\newblock \bibinfo{journal}{\emph{ACM Transactions on Human-Robot Interaction (THRI)}} \bibinfo{volume}{11}, \bibinfo{number}{1} (\bibinfo{year}{2021}), \bibinfo{pages}{1--33}.
\newblock


\bibitem[Do et~al\mbox{.}(2022)]%
        {do2022new}
\bibfield{author}{\bibinfo{person}{Tiffany~D Do}, \bibinfo{person}{Ryan~P McMahan}, {and} \bibinfo{person}{Pamela~J Wisniewski}.} \bibinfo{year}{2022}\natexlab{}.
\newblock \showarticletitle{A new uncanny valley? The effects of speech fidelity and human listener gender on social perceptions of a virtual-human speaker}. In \bibinfo{booktitle}{\emph{Proceedings of the 2022 CHI conference on human factors in computing systems}}. \bibinfo{pages}{1--11}.
\newblock


\bibitem[D’Alonzo et~al\mbox{.}(2019)]%
        {d2019different}
\bibfield{author}{\bibinfo{person}{M D’Alonzo}, \bibinfo{person}{A Mioli}, \bibinfo{person}{D Formica}, \bibinfo{person}{L Vollero}, {and} \bibinfo{person}{G Di~Pino}.} \bibinfo{year}{2019}\natexlab{}.
\newblock \showarticletitle{Different level of virtualization of sight and touch produces the uncanny valley of avatar’s hand embodiment}.
\newblock \bibinfo{journal}{\emph{Scientific reports}} \bibinfo{volume}{9}, \bibinfo{number}{1} (\bibinfo{year}{2019}), \bibinfo{pages}{19030}.
\newblock


\bibitem[Executive Office of~the President and Technology({[n.\,d.]})]%
        {whiteHouse}
\bibfield{author}{\bibinfo{person}{President’s Council of Advisors on~Science Executive Office of~the President} {and} \bibinfo{person}{Technology}.} \bibinfo{year}{[n.\,d.]}\natexlab{}.
\newblock \bibinfo{title}{Supercharging Research: Harnessing Artificial Intelligence to Meet Global Challenges}.
\newblock \bibinfo{howpublished}{https://www.whitehouse.gov}.
\newblock
\newblock
\shownote{Accessed: 2024-07-02}.


\bibitem[Ferrara(2023)]%
        {ferrara2023social}
\bibfield{author}{\bibinfo{person}{Emilio Ferrara}.} \bibinfo{year}{2023}\natexlab{}.
\newblock \showarticletitle{Social bot detection in the age of ChatGPT: Challenges and opportunities}.
\newblock \bibinfo{journal}{\emph{First Monday}} (\bibinfo{year}{2023}).
\newblock


\bibitem[Ferreira Dos~Santos et~al\mbox{.}(2019)]%
        {ferreira2019uncovering}
\bibfield{author}{\bibinfo{person}{Eric Ferreira Dos~Santos}, \bibinfo{person}{Danilo Carvalho}, \bibinfo{person}{Livia Ruback}, {and} \bibinfo{person}{Jonice Oliveira}.} \bibinfo{year}{2019}\natexlab{}.
\newblock \showarticletitle{Uncovering social media bots: a transparency-focused approach}. In \bibinfo{booktitle}{\emph{Companion Proceedings of The 2019 World Wide Web Conference}}. \bibinfo{pages}{545--552}.
\newblock


\bibitem[Gray and Wegner(2012)]%
        {gray2012feeling}
\bibfield{author}{\bibinfo{person}{Kurt Gray} {and} \bibinfo{person}{Daniel~M Wegner}.} \bibinfo{year}{2012}\natexlab{}.
\newblock \showarticletitle{Feeling robots and human zombies: Mind perception and the uncanny valley}.
\newblock \bibinfo{journal}{\emph{Cognition}} \bibinfo{volume}{125}, \bibinfo{number}{1} (\bibinfo{year}{2012}), \bibinfo{pages}{125--130}.
\newblock


\bibitem[Hayati et~al\mbox{.}(2009)]%
        {hayati2009honeyspam}
\bibfield{author}{\bibinfo{person}{Pedram Hayati}, \bibinfo{person}{Kevin Chai}, \bibinfo{person}{Vidyasagar Potdar}, {and} \bibinfo{person}{Alex Talevski}.} \bibinfo{year}{2009}\natexlab{}.
\newblock \showarticletitle{HoneySpam 2.0: Profiling web spambot behaviour}. In \bibinfo{booktitle}{\emph{Principles of Practice in Multi-Agent Systems: 12th International Conference, PRIMA 2009, Nagoya, Japan, December 14-16, 2009. Proceedings 12}}. Springer, \bibinfo{pages}{335--344}.
\newblock


\bibitem[Hayawi et~al\mbox{.}(2023)]%
        {hayawi2023social}
\bibfield{author}{\bibinfo{person}{Kadhim Hayawi}, \bibinfo{person}{Susmita Saha}, \bibinfo{person}{Mohammad~Mehedy Masud}, \bibinfo{person}{Sujith~Samuel Mathew}, {and} \bibinfo{person}{Mohammed Kaosar}.} \bibinfo{year}{2023}\natexlab{}.
\newblock \showarticletitle{Social media bot detection with deep learning methods: a systematic review}.
\newblock \bibinfo{journal}{\emph{Neural Computing and Applications}} \bibinfo{volume}{35}, \bibinfo{number}{12} (\bibinfo{year}{2023}), \bibinfo{pages}{8903--8918}.
\newblock


\bibitem[Himelein-Wachowiak et~al\mbox{.}(2021)]%
        {himelein2021bots}
\bibfield{author}{\bibinfo{person}{McKenzie Himelein-Wachowiak}, \bibinfo{person}{Salvatore Giorgi}, \bibinfo{person}{Amanda Devoto}, \bibinfo{person}{Muhammad Rahman}, \bibinfo{person}{Lyle Ungar}, \bibinfo{person}{H~Andrew Schwartz}, \bibinfo{person}{David~H Epstein}, \bibinfo{person}{Lorenzo Leggio}, {and} \bibinfo{person}{Brenda Curtis}.} \bibinfo{year}{2021}\natexlab{}.
\newblock \showarticletitle{Bots and misinformation spread on social media: Implications for COVID-19}.
\newblock \bibinfo{journal}{\emph{Journal of medical Internet research}} \bibinfo{volume}{23}, \bibinfo{number}{5} (\bibinfo{year}{2021}), \bibinfo{pages}{e26933}.
\newblock


\bibitem[Howard and Kollanyi(2016)]%
        {howard2016bots}
\bibfield{author}{\bibinfo{person}{Philip~N Howard} {and} \bibinfo{person}{Bence Kollanyi}.} \bibinfo{year}{2016}\natexlab{}.
\newblock \showarticletitle{Bots,\# strongerin, and\# brexit: Computational propaganda during the uk-eu referendum}.
\newblock \bibinfo{journal}{\emph{arXiv preprint arXiv:1606.06356}} (\bibinfo{year}{2016}).
\newblock


\bibitem[Ilias and Roussaki(2021)]%
        {ilias2021detecting}
\bibfield{author}{\bibinfo{person}{Loukas Ilias} {and} \bibinfo{person}{Ioanna Roussaki}.} \bibinfo{year}{2021}\natexlab{}.
\newblock \showarticletitle{Detecting malicious activity in Twitter using deep learning techniques}.
\newblock \bibinfo{journal}{\emph{Applied Soft Computing}}  \bibinfo{volume}{107} (\bibinfo{year}{2021}), \bibinfo{pages}{107360}.
\newblock


\bibitem[Imperva(2024)]%
        {imperva2024}
\bibfield{author}{\bibinfo{person}{Imperva}.} \bibinfo{year}{2024}\natexlab{}.
\newblock \bibinfo{booktitle}{\emph{2024 Bad Bot Report}}.
\newblock \bibinfo{type}{{T}echnical {R}eport}. \bibinfo{institution}{Imperva, Inc.}
\newblock


\bibitem[Jalil and Mirza(2009)]%
        {jalil2009review}
\bibfield{author}{\bibinfo{person}{Zunera Jalil} {and} \bibinfo{person}{Anwar~M Mirza}.} \bibinfo{year}{2009}\natexlab{}.
\newblock \showarticletitle{A review of digital watermarking techniques for text documents}. In \bibinfo{booktitle}{\emph{2009 International Conference on Information and Multimedia Technology}}. IEEE, \bibinfo{pages}{230--234}.
\newblock


\bibitem[Jansen(2019)]%
        {jansen2019discovering}
\bibfield{author}{\bibinfo{person}{Dennis Jansen}.} \bibinfo{year}{2019}\natexlab{}.
\newblock \showarticletitle{Discovering the uncanny valley for the sound of a voice}.
\newblock \bibinfo{journal}{\emph{Unpublished master’s thesis]. School of Humanities and Digital Sciences Department of Cognitive Science \& Artificial Intelligence. Tilburg}} (\bibinfo{year}{2019}).
\newblock


\bibitem[K{\"a}tsyri et~al\mbox{.}(2015)]%
        {katsyri2015review}
\bibfield{author}{\bibinfo{person}{Jari K{\"a}tsyri}, \bibinfo{person}{Klaus F{\"o}rger}, \bibinfo{person}{Meeri M{\"a}k{\"a}r{\"a}inen}, {and} \bibinfo{person}{Tapio Takala}.} \bibinfo{year}{2015}\natexlab{}.
\newblock \showarticletitle{A review of empirical evidence on different uncanny valley hypotheses: support for perceptual mismatch as one road to the valley of eeriness}.
\newblock \bibinfo{journal}{\emph{Frontiers in psychology}}  \bibinfo{volume}{6} (\bibinfo{year}{2015}), \bibinfo{pages}{390}.
\newblock


\bibitem[Kolomeets et~al\mbox{.}(2024)]%
        {kolomeets2024experimental}
\bibfield{author}{\bibinfo{person}{Maxim Kolomeets}, \bibinfo{person}{Olga Tushkanova}, \bibinfo{person}{Vasily Desnitsky}, \bibinfo{person}{Lidia Vitkova}, {and} \bibinfo{person}{Andrey Chechulin}.} \bibinfo{year}{2024}\natexlab{}.
\newblock \showarticletitle{Experimental Evaluation: Can Humans Recognise Social Media Bots?}
\newblock \bibinfo{journal}{\emph{Big Data and Cognitive Computing}} \bibinfo{volume}{8}, \bibinfo{number}{3} (\bibinfo{year}{2024}), \bibinfo{pages}{24}.
\newblock


\bibitem[Kudugunta and Ferrara(2018)]%
        {kudugunta2018deep}
\bibfield{author}{\bibinfo{person}{Sneha Kudugunta} {and} \bibinfo{person}{Emilio Ferrara}.} \bibinfo{year}{2018}\natexlab{}.
\newblock \showarticletitle{Deep neural networks for bot detection}.
\newblock \bibinfo{journal}{\emph{Information Sciences}}  \bibinfo{volume}{467} (\bibinfo{year}{2018}), \bibinfo{pages}{312--322}.
\newblock


\bibitem[Kulkarni et~al\mbox{.}(2023)]%
        {kulkarni2023llms}
\bibfield{author}{\bibinfo{person}{Chinmay Kulkarni}, \bibinfo{person}{Tongshuang Wu}, \bibinfo{person}{Kenneth Holstein}, \bibinfo{person}{Q~Vera Liao}, \bibinfo{person}{Min~Kyung Lee}, \bibinfo{person}{Mina Lee}, {and} \bibinfo{person}{Hariharan Subramonyam}.} \bibinfo{year}{2023}\natexlab{}.
\newblock \showarticletitle{LLMs and the Infrastructure of CSCW}. In \bibinfo{booktitle}{\emph{Companion Publication of the 2023 Conference on Computer Supported Cooperative Work and Social Computing}}. \bibinfo{pages}{408--410}.
\newblock


\bibitem[Laurene~Azoulay({[n.\,d.]})]%
        {goldmansachs}
\bibfield{author}{\bibinfo{person}{Brook Dane Dennis~Walsh Laurene~Azoulay, Sung~Cho}.} \bibinfo{year}{[n.\,d.]}\natexlab{}.
\newblock \bibinfo{title}{DUAL DYNAMICS: INVESTING IN AND WITH ARTIFICIAL INTELLIGENCE}.
\newblock \bibinfo{howpublished}{https://www.gsam.com/content/gsam/us/en/individual/market-insights/gsam-insights/perspectives/2024/investing-in-and-with-ai.html}.
\newblock
\newblock
\shownote{Accessed: 2024-07-02}.


\bibitem[McDermott et~al\mbox{.}(2018)]%
        {mcdermott2018botnet}
\bibfield{author}{\bibinfo{person}{Christopher~D McDermott}, \bibinfo{person}{Farzan Majdani}, {and} \bibinfo{person}{Andrei~V Petrovski}.} \bibinfo{year}{2018}\natexlab{}.
\newblock \showarticletitle{Botnet detection in the internet of things using deep learning approaches}. In \bibinfo{booktitle}{\emph{2018 international joint conference on neural networks (IJCNN)}}. IEEE, \bibinfo{pages}{1--8}.
\newblock


\bibitem[Morewedge(2022)]%
        {morewedge2022preference}
\bibfield{author}{\bibinfo{person}{Carey~K Morewedge}.} \bibinfo{year}{2022}\natexlab{}.
\newblock \showarticletitle{Preference for human, not algorithm aversion}.
\newblock \bibinfo{journal}{\emph{Trends in Cognitive Sciences}} \bibinfo{volume}{26}, \bibinfo{number}{10} (\bibinfo{year}{2022}), \bibinfo{pages}{824--826}.
\newblock


\bibitem[Mori and Malik(2003)]%
        {mori2003recognizing}
\bibfield{author}{\bibinfo{person}{Greg Mori} {and} \bibinfo{person}{Jitendra Malik}.} \bibinfo{year}{2003}\natexlab{}.
\newblock \showarticletitle{Recognizing objects in adversarial clutter: Breaking a visual CAPTCHA}. In \bibinfo{booktitle}{\emph{2003 IEEE Computer Society Conference on Computer Vision and Pattern Recognition, 2003. Proceedings.}}, Vol.~\bibinfo{volume}{1}. IEEE, \bibinfo{pages}{I--I}.
\newblock


\bibitem[Mori et~al\mbox{.}(2012)]%
        {mori2012uncanny}
\bibfield{author}{\bibinfo{person}{Masahiro Mori}, \bibinfo{person}{Karl~F MacDorman}, {and} \bibinfo{person}{Norri Kageki}.} \bibinfo{year}{2012}\natexlab{}.
\newblock \showarticletitle{The uncanny valley [from the field]}.
\newblock \bibinfo{journal}{\emph{IEEE Robotics \& automation magazine}} \bibinfo{volume}{19}, \bibinfo{number}{2} (\bibinfo{year}{2012}), \bibinfo{pages}{98--100}.
\newblock


\bibitem[Orabi et~al\mbox{.}(2020)]%
        {orabi2020detection}
\bibfield{author}{\bibinfo{person}{Mariam Orabi}, \bibinfo{person}{Djedjiga Mouheb}, \bibinfo{person}{Zaher Al~Aghbari}, {and} \bibinfo{person}{Ibrahim Kamel}.} \bibinfo{year}{2020}\natexlab{}.
\newblock \showarticletitle{Detection of bots in social media: a systematic review}.
\newblock \bibinfo{journal}{\emph{Information Processing \& Management}} \bibinfo{volume}{57}, \bibinfo{number}{4} (\bibinfo{year}{2020}), \bibinfo{pages}{102250}.
\newblock


\bibitem[Ouni et~al\mbox{.}(2022)]%
        {ouni2022bots}
\bibfield{author}{\bibinfo{person}{Sarra Ouni}, \bibinfo{person}{Fethi Fkih}, {and} \bibinfo{person}{Mohamed~Nazih Omri}.} \bibinfo{year}{2022}\natexlab{}.
\newblock \showarticletitle{Bots and gender detection on Twitter using stylistic features}. In \bibinfo{booktitle}{\emph{International Conference on Computational Collective Intelligence}}. Springer, \bibinfo{pages}{650--660}.
\newblock


\bibitem[Radivojevic et~al\mbox{.}(2024)]%
        {radivojevic2024llms}
\bibfield{author}{\bibinfo{person}{Kristina Radivojevic}, \bibinfo{person}{Nicholas Clark}, {and} \bibinfo{person}{Paul Brenner}.} \bibinfo{year}{2024}\natexlab{}.
\newblock \showarticletitle{LLMs Among Us: Generative AI Participating in Digital Discourse}.
\newblock \bibinfo{journal}{\emph{arXiv preprint arXiv:2402.07940}} (\bibinfo{year}{2024}).
\newblock


\bibitem[Ratajczyk(2022)]%
        {ratajczyk2022shape}
\bibfield{author}{\bibinfo{person}{Dawid Ratajczyk}.} \bibinfo{year}{2022}\natexlab{}.
\newblock \showarticletitle{Shape of the uncanny valley and emotional attitudes toward robots assessed by an analysis of YouTube comments}.
\newblock \bibinfo{journal}{\emph{International Journal of Social Robotics}} \bibinfo{volume}{14}, \bibinfo{number}{8} (\bibinfo{year}{2022}), \bibinfo{pages}{1787--1803}.
\newblock


\bibitem[Rauchfleisch and Kaiser(2020)]%
        {rauchfleisch2020false}
\bibfield{author}{\bibinfo{person}{Adrian Rauchfleisch} {and} \bibinfo{person}{Jonas Kaiser}.} \bibinfo{year}{2020}\natexlab{}.
\newblock \showarticletitle{The false positive problem of automatic bot detection in social science research}.
\newblock \bibinfo{journal}{\emph{PloS one}} \bibinfo{volume}{15}, \bibinfo{number}{10} (\bibinfo{year}{2020}), \bibinfo{pages}{e0241045}.
\newblock


\bibitem[Richard~Harper(2024)]%
        {harper2024ml}
\bibfield{author}{\bibinfo{person}{Dave~Randall Richard~Harper}.} \bibinfo{year}{2024}\natexlab{}.
\newblock \showarticletitle{Machine Learning and the Work of the User}.
\newblock \bibinfo{journal}{\emph{Computer Supported Cooperative Work (CSCW)}} (\bibinfo{year}{2024}).
\newblock


\bibitem[RICHARD~WIKE({[n.\,d.]})]%
        {pew1}
\bibfield{author}{\bibinfo{person}{JANELL FETTEROLF CHRISTINE HUANG SARAH AUSTIN LAURA CLANCY SNEHA~GUBBALA RICHARD~WIKE, LAURA~SILVER}.} \bibinfo{year}{[n.\,d.]}\natexlab{}.
\newblock \bibinfo{title}{Social Media Seen as Mostly Good for Democracy Across Many Nations, But U.S. is a Major Outlier}.
\newblock \bibinfo{howpublished}{https://www.pewresearch.org/global/2022/12/06/social-media-seen-as-mostly-good-for-democracy-across-many-nations-but-u-s-is-a-major-outlier/}.
\newblock
\newblock
\shownote{Accessed: 2024-07-02}.


\bibitem[Savage et~al\mbox{.}(2016)]%
        {savage2016botivist}
\bibfield{author}{\bibinfo{person}{Saiph Savage}, \bibinfo{person}{Andres Monroy-Hernandez}, {and} \bibinfo{person}{Tobias H{\"o}llerer}.} \bibinfo{year}{2016}\natexlab{}.
\newblock \showarticletitle{Botivist: Calling volunteers to action using online bots}. In \bibinfo{booktitle}{\emph{Proceedings of the 19th ACM Conference on Computer-Supported Cooperative Work \& Social Computing}}. \bibinfo{pages}{813--822}.
\newblock


\bibitem[Seering et~al\mbox{.}(2018)]%
        {seering2018social}
\bibfield{author}{\bibinfo{person}{Joseph Seering}, \bibinfo{person}{Juan~Pablo Flores}, \bibinfo{person}{Saiph Savage}, {and} \bibinfo{person}{Jessica Hammer}.} \bibinfo{year}{2018}\natexlab{}.
\newblock \showarticletitle{The social roles of bots: evaluating impact of bots on discussions in online communities}.
\newblock \bibinfo{journal}{\emph{Proceedings of the ACM on Human-Computer Interaction}} \bibinfo{volume}{2}, \bibinfo{number}{CSCW} (\bibinfo{year}{2018}), \bibinfo{pages}{1--29}.
\newblock


\bibitem[Seibt et~al\mbox{.}(2014)]%
        {seibt2014sociable}
\bibfield{author}{\bibinfo{person}{Johanna Seibt}, \bibinfo{person}{Marco N{\o}rskov}, {and} \bibinfo{person}{Raul Hakli}.} \bibinfo{year}{2014}\natexlab{}.
\newblock \bibinfo{booktitle}{\emph{Sociable robots and the future of social relations: Proceedings of Robo-philosophy 2014}}. Vol.~\bibinfo{volume}{273}.
\newblock \bibinfo{publisher}{Ios Press}.
\newblock


\bibitem[Shao et~al\mbox{.}(2017)]%
        {shao2017spread}
\bibfield{author}{\bibinfo{person}{Chengcheng Shao}, \bibinfo{person}{Giovanni~L Ciampaglia}, \bibinfo{person}{Onur Varol}, \bibinfo{person}{Alessandro Flammini}, {and} \bibinfo{person}{Filippo Menczer}.} \bibinfo{year}{2017}\natexlab{}.
\newblock \showarticletitle{The spread of misinformation by social bots}.
\newblock \bibinfo{journal}{\emph{arXiv preprint arXiv:1707.07592}} (\bibinfo{year}{2017}).
\newblock


\bibitem[Stocking and Sumida(2018)]%
        {stocking2018social}
\bibfield{author}{\bibinfo{person}{G Stocking} {and} \bibinfo{person}{N Sumida}.} \bibinfo{year}{2018}\natexlab{}.
\newblock \bibinfo{title}{Social Media Bots Draw Public’s Attention and Concern. Washington: Pew Research Center}.
\newblock
\newblock


\bibitem[Stringhini et~al\mbox{.}(2014)]%
        {stringhini2014harvester}
\bibfield{author}{\bibinfo{person}{Gianluca Stringhini}, \bibinfo{person}{Oliver Hohlfeld}, \bibinfo{person}{Christopher Kruegel}, {and} \bibinfo{person}{Giovanni Vigna}.} \bibinfo{year}{2014}\natexlab{}.
\newblock \showarticletitle{The harvester, the botmaster, and the spammer: On the relations between the different actors in the spam landscape}. In \bibinfo{booktitle}{\emph{Proceedings of the 9th ACM symposium on Information, computer and communications security}}. \bibinfo{pages}{353--364}.
\newblock


\bibitem[Suarez-Lledo and Alvarez-Galvez(2022)]%
        {suarez2022assessing}
\bibfield{author}{\bibinfo{person}{Victor Suarez-Lledo} {and} \bibinfo{person}{Javier Alvarez-Galvez}.} \bibinfo{year}{2022}\natexlab{}.
\newblock \showarticletitle{Assessing the role of social bots during the COVID-19 pandemic: Infodemic, disagreement, and criticism}.
\newblock \bibinfo{journal}{\emph{Journal of Medical Internet Research}} \bibinfo{volume}{24}, \bibinfo{number}{8} (\bibinfo{year}{2022}), \bibinfo{pages}{e36085}.
\newblock


\bibitem[Tang et~al\mbox{.}(2024)]%
        {tang2024science}
\bibfield{author}{\bibinfo{person}{Ruixiang Tang}, \bibinfo{person}{Yu-Neng Chuang}, {and} \bibinfo{person}{Xia Hu}.} \bibinfo{year}{2024}\natexlab{}.
\newblock \showarticletitle{The science of detecting llm-generated text}.
\newblock \bibinfo{journal}{\emph{Commun. ACM}} \bibinfo{volume}{67}, \bibinfo{number}{4} (\bibinfo{year}{2024}), \bibinfo{pages}{50--59}.
\newblock


\bibitem[Valtonen et~al\mbox{.}(2019)]%
        {valtonen2019media}
\bibfield{author}{\bibinfo{person}{Teemu Valtonen}, \bibinfo{person}{Matti Tedre}, \bibinfo{person}{Kati M{\"a}kitalo}, {and} \bibinfo{person}{Henriikka Vartiainen}.} \bibinfo{year}{2019}\natexlab{}.
\newblock \showarticletitle{Media Literacy Education in the Age of Machine Learning.}
\newblock \bibinfo{journal}{\emph{Journal of Media Literacy Education}} \bibinfo{volume}{11}, \bibinfo{number}{2} (\bibinfo{year}{2019}), \bibinfo{pages}{20--36}.
\newblock


\bibitem[Wang et~al\mbox{.}(2015)]%
        {wang2015uncanny}
\bibfield{author}{\bibinfo{person}{Shensheng Wang}, \bibinfo{person}{Scott~O Lilienfeld}, {and} \bibinfo{person}{Philippe Rochat}.} \bibinfo{year}{2015}\natexlab{}.
\newblock \showarticletitle{The uncanny valley: Existence and explanations}.
\newblock \bibinfo{journal}{\emph{Review of General Psychology}} \bibinfo{volume}{19}, \bibinfo{number}{4} (\bibinfo{year}{2015}), \bibinfo{pages}{393--407}.
\newblock


\bibitem[Xu and Sasahara(2022)]%
        {xu2022characterizing}
\bibfield{author}{\bibinfo{person}{Wentao Xu} {and} \bibinfo{person}{Kazutoshi Sasahara}.} \bibinfo{year}{2022}\natexlab{}.
\newblock \showarticletitle{Characterizing the roles of bots on Twitter during the COVID-19 infodemic}.
\newblock \bibinfo{journal}{\emph{Journal of Computational Social Science}} \bibinfo{volume}{5}, \bibinfo{number}{1} (\bibinfo{year}{2022}), \bibinfo{pages}{591--609}.
\newblock


\bibitem[Yang et~al\mbox{.}(2022)]%
        {yang2022botometer}
\bibfield{author}{\bibinfo{person}{Kai-Cheng Yang}, \bibinfo{person}{Emilio Ferrara}, {and} \bibinfo{person}{Filippo Menczer}.} \bibinfo{year}{2022}\natexlab{}.
\newblock \showarticletitle{Botometer 101: Social bot practicum for computational social scientists}.
\newblock \bibinfo{journal}{\emph{Journal of computational social science}} \bibinfo{volume}{5}, \bibinfo{number}{2} (\bibinfo{year}{2022}), \bibinfo{pages}{1511--1528}.
\newblock


\bibitem[Yang and Menczer(2023)]%
        {yang2023anatomy}
\bibfield{author}{\bibinfo{person}{Kai-Cheng Yang} {and} \bibinfo{person}{Filippo Menczer}.} \bibinfo{year}{2023}\natexlab{}.
\newblock \showarticletitle{Anatomy of an AI-powered malicious social botnet}.
\newblock \bibinfo{journal}{\emph{arXiv preprint arXiv:2307.16336}} (\bibinfo{year}{2023}).
\newblock


\bibitem[Zhang et~al\mbox{.}(2016)]%
        {zhang2016rise}
\bibfield{author}{\bibinfo{person}{Jinxue Zhang}, \bibinfo{person}{Rui Zhang}, \bibinfo{person}{Yanchao Zhang}, {and} \bibinfo{person}{Guanhua Yan}.} \bibinfo{year}{2016}\natexlab{}.
\newblock \showarticletitle{The rise of social botnets: Attacks and countermeasures}.
\newblock \bibinfo{journal}{\emph{IEEE Transactions on Dependable and Secure Computing}} \bibinfo{volume}{15}, \bibinfo{number}{6} (\bibinfo{year}{2016}), \bibinfo{pages}{1068--1082}.
\newblock


\bibitem[Zhang and Gosline(2023)]%
        {zhang2023human}
\bibfield{author}{\bibinfo{person}{Yunhao Zhang} {and} \bibinfo{person}{Ren{\'e}e Gosline}.} \bibinfo{year}{2023}\natexlab{}.
\newblock \showarticletitle{Human favoritism, not AI aversion: People’s perceptions (and bias) toward generative AI, human experts, and human--GAI collaboration in persuasive content generation}.
\newblock \bibinfo{journal}{\emph{Judgment and Decision Making}}  \bibinfo{volume}{18} (\bibinfo{year}{2023}), \bibinfo{pages}{e41}.
\newblock


\bibitem[Zhuo et~al\mbox{.}({[n.\,d.]})]%
        {zhuo2023red}
\bibfield{author}{\bibinfo{person}{Terry~Yue Zhuo}, \bibinfo{person}{Yujin Huang}, \bibinfo{person}{Chunyang Chen}, {and} \bibinfo{person}{Zhenchang Xing}.} \bibinfo{year}{[n.\,d.]}\natexlab{}.
\newblock \showarticletitle{Red teaming chatgpt via jailbreaking: Bias, robustness, reliability and toxicity}.
\newblock \bibinfo{journal}{\emph{arXiv preprint arXiv:2301.12867}} (\bibinfo{year}{[n.\,d.]}).
\newblock


\end{thebibliography}

\end{document}